\documentstyle[epsfig]{mn}

\newcommand{\Msolar}{\mbox{\,$\rm M_{\odot}$}}        

\newcommand{\kms}{km~s$^{-1}$}
\begin{document}

\title[]{The blueshift of the [O III] emission line in NLS1s}
\author[Bian Weihao, Yuan Qirong \& Zhao Yongheng]
{W.Bian$^{1,2}$, Q. Yuan$^{1}$ and Y.Zhao$^2$\\
$^{1}$Department of Physics, Nanjing Normal University, Nanjing
210097, China\\
$^{2}$National Astronomical Observatories, Chinese Academy of
Sciences, Beijing 100012, China\\
}
\maketitle
\begin{abstract}
We investigate the radial velocity difference between the narrow
emission-line components of [O III] $\lambda$ 5007 and H$\beta$ in
a sample of 150 SDSS NLS1s.  Seven ``blue outliers" with [O III]
blueshifted by more than 250 \kms are found. A strong correlation
between the [O III] blueshift and the Eddington ratio is found for
these seven ``blue outliers". For the entire sample, we found a
modest correlation between the blueshift and the linewidth of the
narrow component of the [O III] line. The reflected profile of [O
III] indicates two kinematically and physically distinct regions.
The [O III] linewidth depends not only on the bulge stellar
gravitational potential, but also on the central black hole
potential.

\end{abstract}
\begin{keywords}
galaxies:active --- galaxies:nuclei --- quasars: emission lines
\end{keywords}

\section{INTRODUCTION}

As an interesting subclass of active galactic nuclei (AGNs),
narrow-line Seyfert 1 galaxies (NLS1s) were initially defined with
the following optical spectral characteristics: H$\beta$ full
width half maximum (FWHM) less than $2000$ \kms; strong optical Fe
II multiplets; a line-intensity ratio of [O~III]$\lambda$5007 to
H$\beta$ less than 3, which distinguishes between Seyfert 1 and
Seyfert 2 galaxies (Osterbrock \& Pogge 1985; Goodrich 1998). A
steep, soft X-ray excess (Puchnarewicz et al. 1992; Boller et al.
1996) and rapid soft/hard X-ray variability (Leighly 1999; Cheng
et al. 2002) are shown in the X-ray observations of NLS1s. The
popular model of NLS1s is that they contain less massive black
holes with higher Eddington ratios (Pounds et al. 1995; Wandel \&
Boller (1998); Laor et al. 1997; Mineshige et al. 2000),
suggesting that NLS1s might be in the early stage of AGN evolution
(Grupe 1996; Mathur 2000; Bian \& Zhao 2003).

Based on the standard model of AGNs, the characteristic emission
lines are emitted from the so-called broad line region (BLR) and
the narrow line region (NLR) (see Sulentic et al. 2000 for a
review of the BLR). The reverberation mapping technique shows that
the kinematics of the emitting gas in the BLR are determined by
the gravitational potential of the central supermassive black hole
(Peterson \& Wandel 2000). The correlation between the bulge
stellar velocity dispersion (derived from Ca II absorption and [O
III] emission) suggests that the NLR kinematics are determined, in
large part, by the bulge potential (Nelson \& White 1996).
Therefore, the FWHM of the [O III] line may be adopted as a
surrogate of the bulge stellar velocity dispersion. Recently,
using the Sloan Digital Sky Survey (SDSS), we find that NLS1s as a
class deviate from the $M_{\rm BH}-\sigma$ relation seen in normal
galaxies. This is confirmed by other investigations with ROSAT
data, and it is suggested that the black holes with higher
Eddington ratios in some NLS1s are still growing (Grupe \& Mathur
2004; Mathur \& Grupe 2005). Although Boroson (2003) suggested
that for AGNs the stellar velocity dispersions measured by [O III]
can predict the black hole mass to a factor of five, it is
possible that the [O III] width is not a good tracer of the
velocity dispersion for NLS1s. Botte et al. (2005) find that the
[O III] linewidth indeed typically overestimates the stellar
velocity dispersion compared to the direct measure of the Ca II
absorption triplet. These results support that the dynamics of the
NLR in NLS1s would be different than that of other AGNs.

Outflows have been reported in some NLS1s. A very significant
blueshift of about $10\AA$ for the [O III] line relative to the
rest-frame defined by the H$\beta$ line is found in the prototype
NLS1 I ZW 1 (Boroson \& Oke 1987).  This kind of blueshift is
generally interpreted as the NLRs outflow relative to the BLRs.
Zamanov et al. (2002) also find other six AGNs showing the [O III]
blueshifts larger than 250 \kms in their sample of 216 type 1 AGNs
(Marziani et al. 2003a), and these objects are called ``blue
outliers". Two NLS1s with the largest [O III] blueshifts were
recently found by Aoki, Kawaguchi \& Ohta (2005). Up to now, 16
``blue outliers" have been found (Grupe 2001; Verron-Cetty et al.
2001; Grupe \& Leighly 2002; Zamanov et al. 2002; Marziani et al.
2003b; Aoki, Kawaguchi \& Ohta 2005). The distribution of these
``blue outliers" in the eigenvector 1 diagrams is found to be in
the population A region of the Eigenvector 1 parameter domain (see
Fig. 1c in Zamanov et al. 2002). The ``blue outlier" is the
population with strong Fe II lines and narrow H$\beta$ line, very
similar to the characteristics of NLS1s.

In this paper, we use the SDSS NLS1s to find more ``blue outliers".
For the ``blue outliers" in our sample, we carefully model the
emission lines of H$\beta$ and [O III] with multiple components,
which will provide some clues about the NLR dynamics of the
``blue outliers". All of the cosmological calculations in this
paper assume $H_{0}=75 \rm {~km ~s^ {-1}~Mpc^{-1}}$,
$\Omega_{M}=0.3$, $\Omega_{\Lambda} = 0.7$.

\section{SAMPLE AND ANALYSIS}
Using the NLS1 selection criteria outlined above (Osterbrock \&
Pogge 1985; Goodrich 1989), Williams et al. (2003) generated a
sample of 150 NLS1s found within the SDSS Early Data Release
(EDR), which is the largest published sample of NLS1s. The spectra
of these 150 NLS1s are obtained from SDSS Data Release 3 (DR3).
Because of the lack of the [O III] line, SDSS J153243.67-004342.5
is ignored in our analysis.

The spectra are transformed into the rest frame defined by the
redshift given in their FITS header. Because of the
asymmetry of the profiles of [O III] and/or H$\beta$ lines, we
used two-component models to fit the emission lines of H$\beta$
and [O III] $\lambda\lambda$4959, 5007, employing the IRAF task
SPECFIT (Kriss 1994) for a careful look at the [O III] blueshift
phenomenon (e.g. Zheng et al. 2002; Dong et al. 2005). The
Galactic interstellar reddening was corrected using E(B-V)=0.046
(Schlegel et al. 1998), assuming an extinction curve with
$R_V=3.1$.

NLS1s generally show strong Fe II emission lines which contaminate
the continuum and the emission lines of H$\beta$ and
[O III]$\lambda\lambda$4959,5007. The Fe II template, derived from
the prototype NLS1 I ZW 1 (Boroson \& Green 1992), is used to
subtract the Fe II multiples from the spectra. It is broadened by
convolving with a Gaussian of various line widths and scaled by
multiplying by a factor indicating the line strength. At the same
time, a power-law continuum is also fit. The best
Fe II and power-law subtraction is found when the spectral regions between
the H$\gamma$ and H$\beta$ (Fe II multiplets 37, 38) and between
5100\AA and 5400\AA (Fe II multiplets 48, 49) are flat.

We use two Gaussians to model the line profiles of [O III] and
H$\beta$. For the doublet [O III] $\lambda\lambda$4959,5007, we
take the same linewidth for each component, and fix the flux ratio
of [O III]$\lambda$4959 to [O III]$\lambda$5007 to be 1:3.

We measure the wavelength difference between the centroid of the
narrow components of H$\beta\lambda$4861.3 and [O~III]$\lambda$5006.8
lines (i.e. the peaks of the H$\beta$ and [O III] lines).
Then the [O III] blueshift relative to H$\beta$ (i.e.
$\Delta V$) in units of \kms can be calculated based on the
the laboratory wavelength difference (145.5\AA).  In order
to investigate the drivers of the [O III] blueshift, we also
calculated the Eddington ratio, i.e. the bolometric luminosity as
a fraction of the Eddington luminosity ($L_{bol}/L_{Edd}$). We
take $L_{bol}=9\lambda L_{\lambda}(5100\AA)$ (Kaspi et al. 2000)
and $L_{Edd}=1.26\times 10^{38}M_{bh}/\Msolar~ergs~ s^{-1}$, where
$M_{bh}$ is the black hole mass. $\lambda L_{\lambda}(5100\AA)$ is
derived from the r$^\ast$ magnitude. $M_{bh}$ is calculated from
H$\beta$ FWHM and the empirical size-luminosity formula (Kaspi et
al. 2000). $\lambda L_{\lambda}(5100\AA)$ and $M_{bh}$ values are
respectively taken from Col.4 and Col.5 in Table 1 of Bian \& Zhao
(2004a). The best-fit Fe II flux between 4434$\AA$ and 4684$\AA$
is calculated from Fe II spectra and the flux ratio of the Fe II
(between 4434 and 4684\AA) to H$\beta$ (Fe II$\lambda$ 4570/H$\beta$)
is also calculated.

Following Zamanov et al. (2002), we take $\Delta V < -250$ \kms as
the criterion for classification as a ``blue outlier" and finally
select the blue outliers in the sample of SDSS NLS1s.

\section{RESULTS}

\subsection{Distributions of the blueshift and the Eddington ratio}

Fig. 1 shows the distribution of the derived [O III] blueshifts
for these 149 NLS1s. The mean value of the [O III] blueshift is
$-52\pm 13$ \kms with a standard deviation of 158 \kms.

It is generally accepted that the Eddington ratio is large in
NLS1s, which is possibly related to the outflow phenomena. Fig. 1
also shows the distribution of the Eddington ratios in the logarithm.
The mean value is $-0.06\pm 0.02$ with a standard deviation of 0.26.
Many NLS1s have relatively small Eddington ratios. Considering the
correlation between the X-ray spectral index and the Eddington ratio
(Grupe 2004; Bian 2005), these objects with lower Eddington ratios
should not have very steep X-ray photo indices, which is consistent with
recent Chandra observations of some SDSS NLS1s
(Williams, Mathur \& Pogge 2004).

\subsection{Correlations between the blueshift and other parameters}

In Fig. 2, we show the [O III] blueshift as a function of the
black hole mass, the Eddington ratio, and Fe II $\lambda$4570/H$\beta$.
Using the simple least
square linear regression (Press et al. 1992), no significant correlations
are found. We also plot in Fig. 2 the [O III] blueshift versus FWHM of
the narrow component of the [O III] line. The simple least
square linear regression shows a modest correlation with a correlation
coefficient is 0.54. NLS1s with larger [O III] blueshift tend to show
broader [O III] linewidth.

\subsection{The ``blue outliers"}

Although the spectra have low signal-to-noise ratios, larger
displacements in some NLS1s exist in both [O III]$\lambda$4959 and
[O III]$\lambda$5007 relative to the rest frame defined by the
narrow component of H$\beta$ line. The errors and uncertainties
are discussed in Sec. 4.1.

Taking $\Delta V < -250$ \kms as our criterion (Zamanov et al.
2002), we found 13 NLS1s to be ``blue outliers.'' For some NLS1s
with low signal-to-noise spectra, the wavelength error of the line
centroid (for the narrow/broad components of H$\beta$/[O III]
lines) in the fitting is sometimes large. Because the line
centroid error of some component is larger than 5$\AA$, we
consider six objects in these 13 NLS1s to be questionable
classifications. At last we obtain seven reliable ``blue
outliers." The parameters of these seven ``blue outliers" are
listed in Table 1. Considering the lower limit of the blueshift,
one NLS1 does not satisfy the criterion of $\Delta V < -250$ \kms,
and that is SDSS J115533.50+010730.6.

In Fig. 2, red squares denote the seven reliable ``blue outliers".
Using the simple least square linear regression (Press et al. 1992),
correlations between the [O III] blueshift and the
black hole mass, the Eddington ratio, Fe II $\lambda$4570/H$\beta$
are found and the correlation coefficients are listed in Table 2.
``Blue outliers" tend to show broad [O III] linewidth.

Fig. 3 shows rest-frame spectra for the ``blue outliers" (left).
The Fe II spectra are shown in the bottom in each
left panel. Remarkable Fe II emission is obvious. The value of Fe
II$\lambda$ 4570/H$\beta$ is listed in Col. 9 in Table 1. We also
plot two-component fitting of H$\beta$ and [O III]$\lambda\lambda$4959, 5007
(right) in Fig. 3. The fitting depends on the spectra quality. The fitting
goodness is illustrated by the residual spectrum shown in the bottom in each
right panel. Using different initial-values, the best values and the
errors for different line components (e.g. FWHM, the line flux,
and the [O III] blueshift relative to the narrow component of
H$\beta$) are listed in Table 3 for these ``blue outliers".

\section{DISCUSSION}
\subsection{Errors}

We have found obvious differences between the spectra of the same
NLS1s released from SDSS EDR and DR3, up to a factor of 1.5 for
some NLS1s. This would lead to the uncertainty of 0.1 dex at most
in the mass estimate. In Fig. 3 of Bian \& Zhao (2004a), we found
that monochromatic luminosity at 5100$\AA$ measured from the
spectra of the SDSS NLS1s is lower than that derived from
$r^{\ast}$ band magnitude by 0.21 in the logarithm scale. The new
flux calibration shows that the monochromatic luminosity at
5100$\AA$ measured from the spectra is consistent with that
derived from $r^{\ast}$ band magnitude. The optical magnitude
(here we adopted the $r^{\ast}$ band magnitude) is popularly used to
derive the BLR size and then to calculate the black hole virial
mass from the H$\beta$ FWHM (e.g. Wang \& Lu, 2001; Gu, Cao \&
Jiang, 2001; Bian \& Zhao 2004a; Bian \& Zhao 2004b). Using this
method, we can estimate the black hole virial mass within a factor
of 3 (i.e., $\sim 0.5$ dex) (Bian \& Zhao 2004a, Bian \& Zhao
2004b). The error in the Eddington ratio, ($L_{bol}/L_{Edd}$)
largely depends on the error in the black hole mass. They are
almost the same, i.e. $\sim 0.5$ dex.

The [O III] blueshift relative to H$\beta$ is calculated from their
centroid wavelength difference. The error of the [O III] blueshift
depends on the errors of the centroid wavelengthes of the [O III]
line and H$\beta$ line, and is calculated from propogating their
errors, which are shown in Fig. 1. For our ``blue outliers", the errors
of the blueshifts are also listed in Table 3. The typical errors of the
centroid of the narrow component of the [O III] line and the
H$\beta$ line are 1 \AA, i. e., 60\kms. For the linewidth and the
line flux, the typical error is about 10\%. However, the
systematic effects are neglected, e.g., the uncertainties of the
continuum substraction, the Fe II template, and component
decomposition (Greene \& Ho 2005). The instrumental FWHM is about
60 \kms for the [O III] line, which is small relative to the [O
III] FWHM. Therefore, the centroid of the narrow components of
H$\beta$ and [O III] is reliable.

\subsection{Largest [O III] blueshift: SDSS J010226.31-003904.6}
The ``blue outlier" with the largest [O III] blueshift
is SDSS J010226.31-003904.6 (see Table 1). It has
the strongest Fe II$\lambda$4570/H$\beta$ as well.
Strong Fe II emission lines contaminate its continuum
and seriously blend with the lines of H$\beta$ and [O
III]$\lambda\lambda$4959,5007. There are two peaks in the region
of [O III]$\lambda$5007.  One peak is due to Fe II emission
(see Fig. 3). After the Fe II template is subtracted, we found
that the profile of [O III]$\lambda$5007 is asymmetric and
used a two-component fitting (see Fig. 3). Although the spectrum
is noisy, we still found that the blueshift of the narrow
component of [O III] relative the narrow component of H$\beta$ is
$-931\pm 46$ \kms. Aoki et al. (2005) also presented this object
as one of the two largest blueshifts and they measured the blueshift
to be $-880\pm 30$ \kms. These two measurements are consistent
within their errors. $M_{bh}$ is $10^{7.86} \Msolar$ and $L_{bol}/L_{Edd}$
is 0.034, which are also consistent with the results from Aoki et al. (2005)
considering the uncertainties of $\sim 0.5$ dex in $M_{bh}$ and
$L_{bol}/L_{Edd}$. Other parameters are listed in the first line
in Table 1.

\subsection{Drivers of the [O III] blueshift}

It is clear that [O III] blueshifts relative to H$\beta$ exist
in AGN spectra, especially in NLS1s (see Fig. 3 and Table 3.).
The distribution of the [O III] blueshifts is very
similar to the result of Zamanov et al. (2002) (see Fig. 1b
therein). The origin of the [O III] blueshift relative to H$\beta$
is still a question of debate. The standard model of AGNs is the
disk accretion around the central supermassive black hole with the
BLR, NLR, and jet. The [O III] blueshift is generally
interpreted as an indicator of the outflow in AGNs (e.g. Aoki, et
al. 2005). Marziani et al. (2003b) find that ``blue outliers" have
higher Eddington ratios, and the accretion may be the engine of
this outflow.

However, no correlation is found between the [O III] blueshift and
the Eddington ratio in another sample of 16 ``blue outliers"
(Aoki et al. 2005). They suggested that the [O III]
blueshift only weakly depends on the Eddington ratio. Here we also
showed the relations between the [O III] blueshift and the
Eddington ratio, black hole mass in Fig. 2. Although no
correlation is found for the entire NLS1 sample, a strong
correlation is found for our ``blue outliers." The relationship
between the [O III] blueshift and $L_{bol}/L_{Edd}$ is stronger
than that between [O III] blueshift and $M_{bh}$ (see Table 2).
However, no correlation is found if we combine the Aoki et al. (2005)
16 ``blue outliers" and ours together. This is possibly due to
the larger uncertainties in the Eddington ratio.

The blueshift of [O III] is possibly the result of the outflowing
gas from the nucleus and the obscuration of the receding part of the
flow which depends on the viewing angle (Zamanov et al. 2002). The
outflowing gas is possibly from the inner NLR, which is related
to the wind of the accretion disk and would have larger linewidth
(Elvis 2000). The blueshift and larger linewidth of [O III] can be
interpreted in this scenario (Zamanov et al. 2002).

Some possible necessary conditions are suggested to produce the
outflow (Aoki et al. 2005): larger black hole masses
($>10^{7}\Msolar$), higher accretion rates ($> 2\Msolar~yr^{-1}$),
or larger luminosity ($\lambda
L_{\lambda}(5100\AA)>10^{44.6}~ergs~s^{-1}$) (see Fig. 8 therein).
Based on our ``blue outliers" found in SDSS sample (see Table 1),
the black hole masses in most of our ``blue outliers" are
larger($>10^{7}\Msolar$). The condition of larger luminosity is
not confirmed by our sample.

Aoki et al. (2005) suggested there is a strong correlation between
the [O III] blueshift and the radio luminosity for their
16 ``blue outliers". Here we find other six ``blue outliers" in
SDSS NLS1s. Radio data for these ``blue outliers" is needed to
check this correlation. For radio-loud NLS1s, we should used the
H$\beta$ luminosity instead of $\lambda L_{\lambda}(5100\AA)$ to
derive the black hole mass (Wu et al. 2004; Kaspi et al. 2005).

\subsection{Dynamics of [O III]}
The [O III] FWHM or its narrow/core component can be used as an
indicator of the black hole mass (Nelson \& Whittle 1996; Greene
\& Ho 2005). Our ``blue outliers" tend to show large [O III] FWHM,
which is consistent with the results of Aoki et al. (2005). For
all SDSS NLS1s in our sample, the correlation coefficient is -0.55
for the relation between the [O III] blueshift and the FWHM of the
narrow component of [O III] line. However, for our ``blue
outliers," this relation is very weak. We also used the [O III]
FWHM measured with single Gaussian fit ($FWHM^{one} ([O
III]$), Col. 6 in Table 1 in Bian \& Zhao 2004a). However no
correlation is found between the [O III] blueshift and $FWHM^{one}
([O III])$ for all SDSS NLS1s or the ``blue outliers". Combining the
Aoki et al.  16 ``blue outliers" and our ``blue outliers" together, a modest
correlation is found between the [O III] blueshift and $FWHM^{one}
([O III])$  is found (R=0.54, P=0.007, Fig. 4).

The two components of the [O III]$\lambda \lambda$ 4959, 5007
lines are found to have different linewidths and blueshifts
relative to the narrow H$\beta$ line. We interpret these two
components as two kinematically and physically distinct regions in
this blue outlier (Holt, Tadhunter \& Morganti, 2003). The broad
component of [O III]$\lambda$5007 is almost the same order of magnitude
as expected for the BLR.  This is consistent with the above
outflow/wind scenario for the inner NLR, which is influenced by
the central black hole potential. Therefore, the [O III] complex
indicates that its width depends not only on the bulge stellar
gravitational potential, but also on the central black hole
potential.

Verron-Cetty (2001) suggests that Lorentzian profiles rather than
Gaussians are more suitable to reproduce the shape of NLS1 broad
emission lines. Considering the low signal-to-noise ratio of the SDSS
NLS1 spectra, we found Gaussian profiles work fine. It is
normally expected that the line ratio of [O III]$\lambda$5007 to
H$\beta$ is about 10, and such a blueshift of the narrow component
of H$\beta$ line from NLRs will be lost in the noise.
Rodriguez-Ardila et al. (2000) suggested that the [O
III]$\lambda$5007/H$\beta$ ratio emitted in NLR varies from 1 to
5, instead of the normally expected value of 10. If that is the
case, we should use FWHM of the broad component of H$\beta$ to
calculate the black hole mass, and we should pay more attention
to our black hole mass estimate and the Eddington ratio. We found
that the NLR contribution needs to subtracted to calculate the black
hole mass correctly, especially for objects with larger
[O III] flux. This will presented in our next paper. For our ``blue
outliers," the [O III] flux is relatively weak. Therefore, as far
as the H$\beta$ emission line is concerned, the contribution from
NLR in these ``blue outliers" can be neglected.

\section{CONCLUSIONS}
The blueshift of the narrow component of [O III] relative to the
narrow component of H$\beta$ is investigated for a sample of 150
SDSS NLS1s. The main conclusions can be summarized as follows.

\begin{itemize}
\item{We have measured the radial velocity difference between the
narrow components of [O III] $\lambda$5007 and H$\beta$ lines for
149 SDSS NLS1s. The mean value of the [O III] blueshift relative
to H$\beta$ is $-52\pm 13$ \kms with a standard deviation of 158
\kms.}

\item{We have found seven ``blue outliers" ($\sim 5\%$ in the SDSS
NLS1 sample) if taking $250 km~ s^{-1}$ (i.e. about 4\AA) as the
criterion of the ``blue outlier." Considering one lower limit,
six NLS1s satisfied this criterion. The ``blue
outliers" with the largest [O III] blueshift is SDSS
J010226.31-003904.6 with a measurement of $-931\pm 46$,
which is consistent with the results of Aoki et al. (2005). }

\item{Strong correlations are found between the amount of the [O
III] blueshift and the Eddington ratio, and Fe II $\lambda
4570$/H$\beta$, which supports the outflow/wind scenario
as the origin of the [O III] blueshift. The correlation of $\Delta V -
L_{bol}/L_{Edd}$ is stronger than the correlation of $\Delta V -
M_{bh}$. For the whole sample of 149 NLS1s, we also found a modest
correlation between the [O III] blueshift and FWHM of the narrow
component of the [O III] line.}

\end{itemize}

\section*{ACKNOWLEDGMENTS}
We thank Dr. X. Zheng for helpful discussions. We thank the
referee, Dr. D. Grupe, for his useful remarks. We are grateful to
Dr. Helmut Abt and Dr. M. Brotherton for checking our manuscript.
This work has been supported by the NSFC (No. 10403005; No.
10473005; No. 10273007) and the NSF from Jiangsu Provincial
Education Department (No. 03KJB160060). Funding for the creation
and distribution of the SDSS Archive has been provided by the
Alfred P. Sloan Foundation, the Participating Institutions, NASA,
the National Science Foundation, the US Department of Energy, the
Japanese Monbukagakusho, and the Max Planck Society. The SDSS Web
site is http:// www.sdss.org/. This research has made use of the
NASA/IPAC Extragalactic Database, which is operated by the Jet
Propulsion Laboratory at Caltech, under contract with the NASA.

\begin{table*}
\begin{small}
\begin{tabular}{llcccccccc}
\hline \hline

Name & z & FWHM of H$\beta$ & FWHM of [O III]&
$\lambda L_{\lambda}$ & $log M_{bh}$& $log L_{bol}/L_{Edd}$& $\Delta V ($\kms$)$ & FeII/H$\beta$\\
(1) & (2) &(3) & (4) & (5) & (6)&(7)& (8) & (9)\\

 \hline

SDSS J010226.31-003904.6 &  0.294  &  1680 $\pm$ 60 & 647$\pm$120  &45.04 & 7.86 & 0.03387  & -931  $\pm$ 46 & 1.646 $\pm$ 0.037 \\
SDSS J013521.68-004402.2 &  0.098  &  1181$\pm$ 182 & 711$\pm$85  &43.45 & 6.44 & -0.13613 & -576  $\pm$ 47 & 1.123 $\pm$ 0.185 \\
SDSS J015249.76+002314.7 &  0.589  &  1852$\pm$110 & 1277$\pm$165 &44.71 & 7.72 & -0.15613 & -501  $\pm$ 78 & 1.333 $\pm$ 0.054 \\
SDSS J024037.89+001118.9 &  0.47   &  1789$\pm$115 & 381$\pm$ 70 &44.19 & 7.32 & -0.27613 & -458  $\pm$ 78 & 1.276 $\pm$ 0.092\\
SDSS J101314.86-005233.5 &  0.276  &  1578$\pm$ 87 & 991$\pm$115  &44.29 & 7.28 & -0.13613 & -314   $\pm$ 31 & 0.866 $\pm$ 0.039\\
SDSS J115533.50+010730.6 &  0.197  &  1628$\pm$202 & 919$\pm$85  &43.76 & 6.94 & -0.32613 & -283  $\pm$ 71 & 0.844 $\pm$0.12\\
SDSS J143230.99-005228.9 &  0.362  &  1559$\pm$121 & 1222$\pm$92 &43.99 & 7.06 & -0.21613 & -288  $\pm$ 31 & 0.802 $\pm$ 0.07\\
\hline
\end{tabular}
\end{small}
\caption{Seven ``blue outliers" found in our sample of 150 SDSS
NLS1s. Columns are: (1) name; (2) redshift; (3) FWHM of H$\beta$
(one gaussian fitting, Williams, Pogge \& Mathur, 2002); (4) the
[O III] FWHM (one-Gaussian fitting, Bian \& Zhao 2004a); (5) log
of continuum luminosity at the rest wavelength 5100 \AA\ in unit
of $erg s^{-1}$; (6) log of the black hole mass in units of solar
mass; (7) the Eddington ratio; (8) the [O III] blueshift relative
to the narrow component of H$\beta$ in units of \kms ; (9) the
flux ratio of the Fe II (between 4434 and 4684 \AA) to H$\beta$. }
\end{table*}

\begin{table*}
\begin{small}
\begin{tabular}{lcccccccc}
\hline \hline

X   & R & SD &
P  & a &b \\
(1) & (2) &(3) & (4) & (5) & (6) \\

 \hline

$Log M_{bh}$  & -0.43 & 5.02 & 0.33    &$1288.4\pm 306.2$& $-240.2\pm42.5$ \\
$Log L_{bol}/L_{Edd}$  & -0.81 & 3.40 & 0.03    &$-736.7\pm 33.0$& $-1937.6\pm183.9$ \\
$Fe II \lambda 4570/H\beta$  & -0.96 & 1.53 & $<10^{-4}$  &$308.0\pm 61.8$& $-726.7\pm58.0$ \\
$FWHM^{narrow}[O III]$ $^\ast$ & -0.55 & 3.76 & $<10^{-4}$    &$133.1\pm 5.5$& $-0.50\pm0.02$ \\
\hline
\end{tabular}
\end{small}
\caption{Correlations between the blueshift and some parameters
for our ``blue outliers". $^\ast$ shows that it is done for all
SDSS NLS1s.  Col. 1: other parameters; Col. 2: correlation
coefficient; Col. 3: standard deviation; Col. 4: probability that
the correlation is caused by a random factor; Col. 5, 6: a and b
in $\Delta V$=a+bX. }
\end{table*}

\begin{table*}
\begin{small}
\begin{tabular}{lcccccccc}
\hline
 \hline

Line & Component & Rest Wavelength &
Velocity shift  & FWHM &Line flux \\
& &  ($\AA$)&(\kms) &(\kms) &($10^{-16}
erg s^{-1} cm^{-2}$)\\
(1) & (2) &(3) & (4) & (5) & (6) \\
\hline
\multicolumn{6}{c}{SDSS J010226.31-003904.6}\\
\hline

{H$\beta$}               & n& $4861.3\pm$ 0.2    & 0    &$1111\pm 46$& $95.6\pm5.0$ \\
                         & b& $4852.2\pm$ 0.6     & $-559\pm 39$ &$4087\pm 97$& $224.4\pm5.2$ \\
{[O III]$\lambda$4959}   & n& $4943.4\pm$ 0.7     & $-944\pm 46$ &$703\pm 101$&  $4.3\pm 1.1$ \\
                         & b& $4930.2\pm$ 2.4     & $-1742\pm 146$ &$1502\pm 176$&  $6.0\pm 3.3$ \\
{[O III]$\lambda$5007}   & n& $4991.3\pm$ 0.7     & $-931\pm 46$  &$703\pm 101$& $12.9\pm 3.3$ \\
                         & b& $4977.9\pm$ 2.4     & $-1730\pm 146$ &$1502\pm 176$& $18.0\pm 9.9$ \\
\hline

\multicolumn{6}{c}{SDSS J013521.68-004402.2}\\
\hline

{H$\beta$}               & n& $4861.3\pm$ 0.8    & 0    &$1269\pm 162$& $19.0\pm4.1$ \\
                         & b& $4848.9\pm$ 4.4     & $-766\pm 277$ &$3413\pm 403$& $22.9\pm4.3$ \\
{[O III]$\lambda$4959}   & n& $4949.3\pm$ 0.2     & $-578\pm 47$ &$617\pm 28$&  $12.3\pm 1.1$ \\
                         & b& $4940.6\pm$ 1.9     & $-1104\pm 125$ &$1082\pm 142$&  $4.3\pm 1.0$ \\
{[O III]$\lambda$5007}   & n& $4997.2\pm$ 0.2     & $-576\pm 47$  &$617\pm 28$& $37.0\pm 3.3$ \\
                         & b& $4988.4\pm$ 1.9     & $-1087\pm 125$ &$1082\pm 143$& $12.9\pm 3.0$ \\
\hline

\multicolumn{6}{c}{SDSS J015249.76+002314.7}\\
\hline

{H$\beta$}               & n& $4861.3\pm$ 0.3    & 0    &$1128\pm 75$& $12.4\pm1.2$ \\
                         & b& $4857.1\pm$ 0.8     & $-258\pm 55$ &$4177\pm 182$& $31.8\pm1.1$ \\
{[O III]$\lambda$4959}   & n& $4950.5\pm$ 1.3     & $-503\pm 78$ &$653\pm 63$&  $0.5\pm 0.4$ \\
                         & b& $4943.1\pm$ 3.5     & $-953\pm 211$ &$1076\pm 195$&  $1.0\pm 0.4$ \\
{[O III]$\lambda$5007}   & n& $4998.4\pm$ 1.3     & $-501\pm 78$  &$653\pm 63$& $1.6\pm 1.2$ \\
                         & b& $4990.9\pm$ 3.5     & $-936\pm 211$ &$1076\pm 195$& $2.9\pm 1.2$ \\
\hline

\multicolumn{6}{c}{SDSS J024037.89+001118.9}\\
\hline

{H$\beta$}               & n& $4861.3\pm$ 0.4    & 0    &$1252\pm 91$& $10.2\pm1.5$ \\
                         & b& $4860.3\pm$ 0.9     & $-64\pm 59$ &$3270\pm 204$& $18.1\pm1.4$ \\
{[O III]$\lambda$4959}   & n& $4951.2\pm$ 1.2     & $-465\pm 78$ &$1118\pm 116$&  $1.8\pm 0.4$ \\
                         & b& $4935.7\pm$ 4.6     & $-1404\pm 277$ &$1867\pm 400$&  $1.1\pm 0.4$ \\
{[O III]$\lambda$5007}   & n& $4999.2\pm$ 1.2     & $-458\pm 78$  &$1118\pm 116$& $5.5\pm 1.2$ \\
                         & b& $4983.5\pm$ 4.6     & $-1390\pm 277$ &$1867\pm 400$& $3.3\pm 1.2$ \\
\hline

\multicolumn{6}{c}{SDSS J101314.86-005233.5}\\
\hline

{H$\beta$}               & n& $4861.3\pm$ 0.3    & 0    &$856\pm 67$& $22.0\pm2.4$ \\
                         & b& $4859.6\pm$ 0.4     & $-104\pm 31$ &$2976\pm 100$& $71.6\pm2.4$ \\
{[O III]$\lambda$4959}   & n& $4953.6\pm$ 0.4     & $-321\pm 31$ &$723\pm 100$&  $4.5\pm 1.6$ \\
                         & b& $4947.0\pm$ 4.2    & $-720\pm253 $&$1313\pm 211$&  $3.2\pm 1.7$ \\
{[O III]$\lambda$5007}   & n& $5001.6\pm$ 0.4     & $-314\pm 31$  &$723\pm 100$& $13.6\pm 4.8$ \\
                         & b& $4994.9\pm$ 4.2     & $-706\pm 253$ &$1313\pm 211$& $9.5\pm 5.1$ \\
\hline

\multicolumn{6}{c}{SDSS J115533.50+010730.6}\\
\hline

{H$\beta$}               & n& $4861.3\pm$ 0.6    & 0    &$996\pm 191$& $13.8\pm4.9$ \\
                         & b& $4858.5\pm$ 1.2     & $-175\pm 79$ &$2595\pm 267$& $32.2\pm4.7$ \\
{[O III]$\lambda$4959}   & n& $4954.1\pm$ 1.0     & $-291\pm 71$ &$736\pm 36$&  $2.5\pm 0.8$ \\
                         & b& $4950.1\pm$ 2.0    & $-533\pm126 $&$894\pm 82$&  $1.5\pm 0.8$ \\
{[O III]$\lambda$5007}   & n& $5002.1\pm$ 1.0      & $-283\pm71$ &$736\pm 36$&  $7.4\pm 2.4$ \\
                         & b& $4998.0\pm$ 2.0     & $-519\pm 126$  &$894\pm 82$& $4.5\pm 2.4$ \\
\hline

\multicolumn{6}{c}{SDSS J143230.99-005228.9}\\
\hline

{H$\beta$}               & n& $4861.3\pm$ 0.4    & 0    &$999\pm 86$& $6.6\pm0.8$ \\
                         & b& $4866.5\pm$ 1.2     & $-321\pm 76$ &$3487\pm 218$& $14.2\pm0.9$ \\
{[O III]$\lambda$4959}   & n& $4954.0\pm$ 0.4     & $-297\pm 31$ &$685\pm 71$&  $1.7\pm 0.3$ \\
                         & b& $4942.2\pm$ 1.7    & $-1015\pm101 $&$1595\pm 97$&  $2.7\pm 0.3$ \\
{[O III]$\lambda$5007}   & n& $5002.0\pm$ 0.4      & $-288\pm31 $&$685\pm 71$&  $5.0\pm 0.9$ \\
                         & b& $4990.0\pm$ 1.7     & $-1002\pm 101$  &$1595\pm 97$& $0.1\pm 0.3$ \\
\hline

\end{tabular}
\end{small}
\caption{Results of the multi-component profile fitting of SDSS
J143230.99-005228.9. Columns are: (1) emission line; (2) emission
line components, where ``n'' and ``b'' represent the narrow and
broad components, respectively; (3) the rest wavelength of the
line component relative to narrow H$\beta$ component in angstroms;
(4) velocity shift of each component relative to the narrow
H$\beta$ component for a particular emission line (\kms); (5) FWHM
of the line components (\kms); (6) integrated line flux ($10^{-16}
erg s^{-1} cm^{-2}$). }
\end{table*}

\newpage

\begin{figure*}
\begin{center}
\epsfig{figure=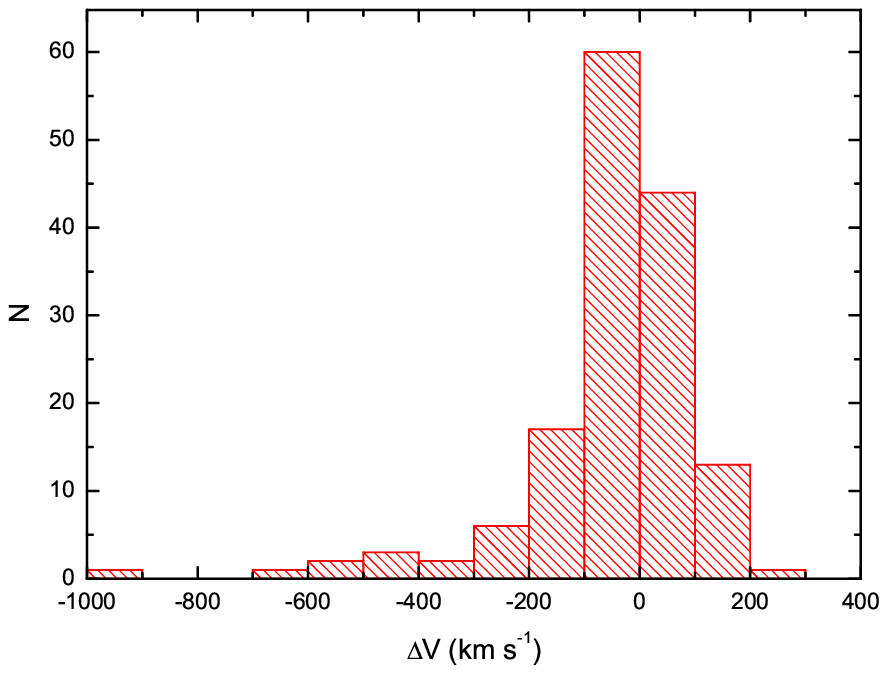,width=8cm}
\epsfig{figure=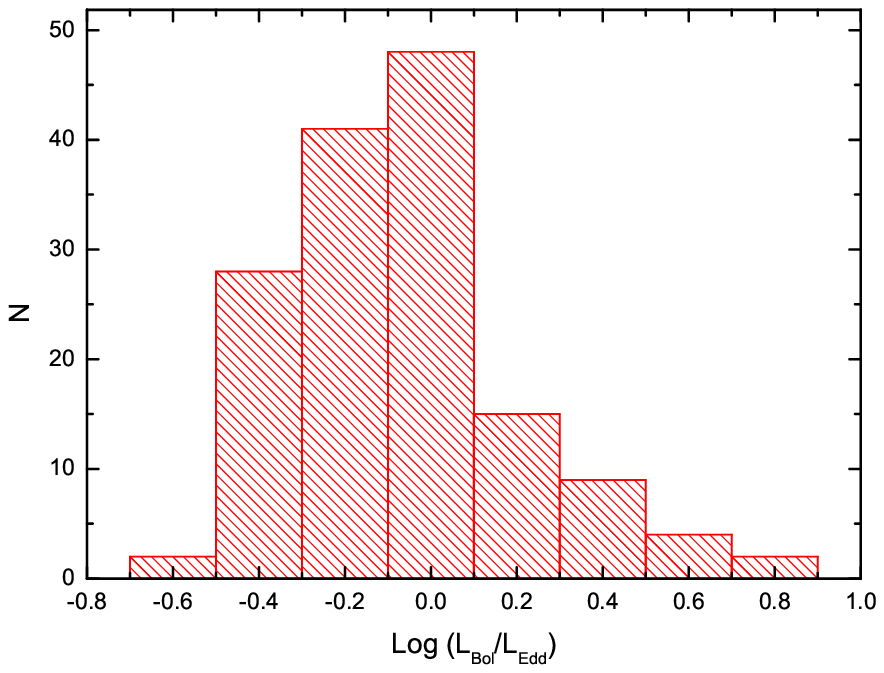,width=8cm}
\caption{The distribution of
the [O III] blueshift relative to H$\beta$ ($\Delta~V$) in units of
\kms (left) and the distribution of $L_{bol}/l_{Edd}$ (right) for
SDSS NLS1s.}
\end{center}
\end{figure*}

\begin{figure*}
\begin{center}
\vglue-3.5cm \epsfig{figure=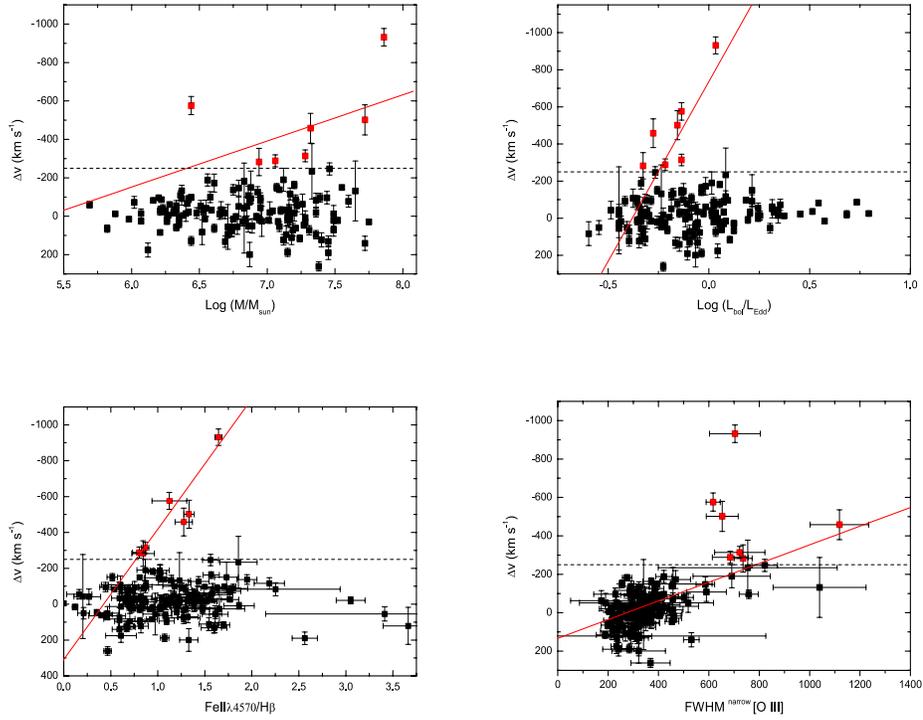,width=17cm,height=25cm}
\vglue-10.5cm

\caption{The [O III] blueshift versus $M_{bh}$, $L_{bol}/l_{Edd}$,
$Fe II\lambda 4570/H\beta$, and $FWHM^{narrow}([O III])$. The red
squares denote the ``blue outliers". The dash line shows -250
\kms. The red solid lines in the figures of $\Delta - M_{bh},
L_{bol}/l_{Edd}, Fe II\lambda 4570/H\beta$ are our best fit for
seven ``blue outliers." The red solid lines in the figure of
$\Delta - FWHM^{narrow}([O III])$ is the best fit for all 149 SDSS
NLS1s. }
\end{center}
\end{figure*}

\begin{figure*}
\begin{center}
\vglue-1.5cm \epsfig{figure=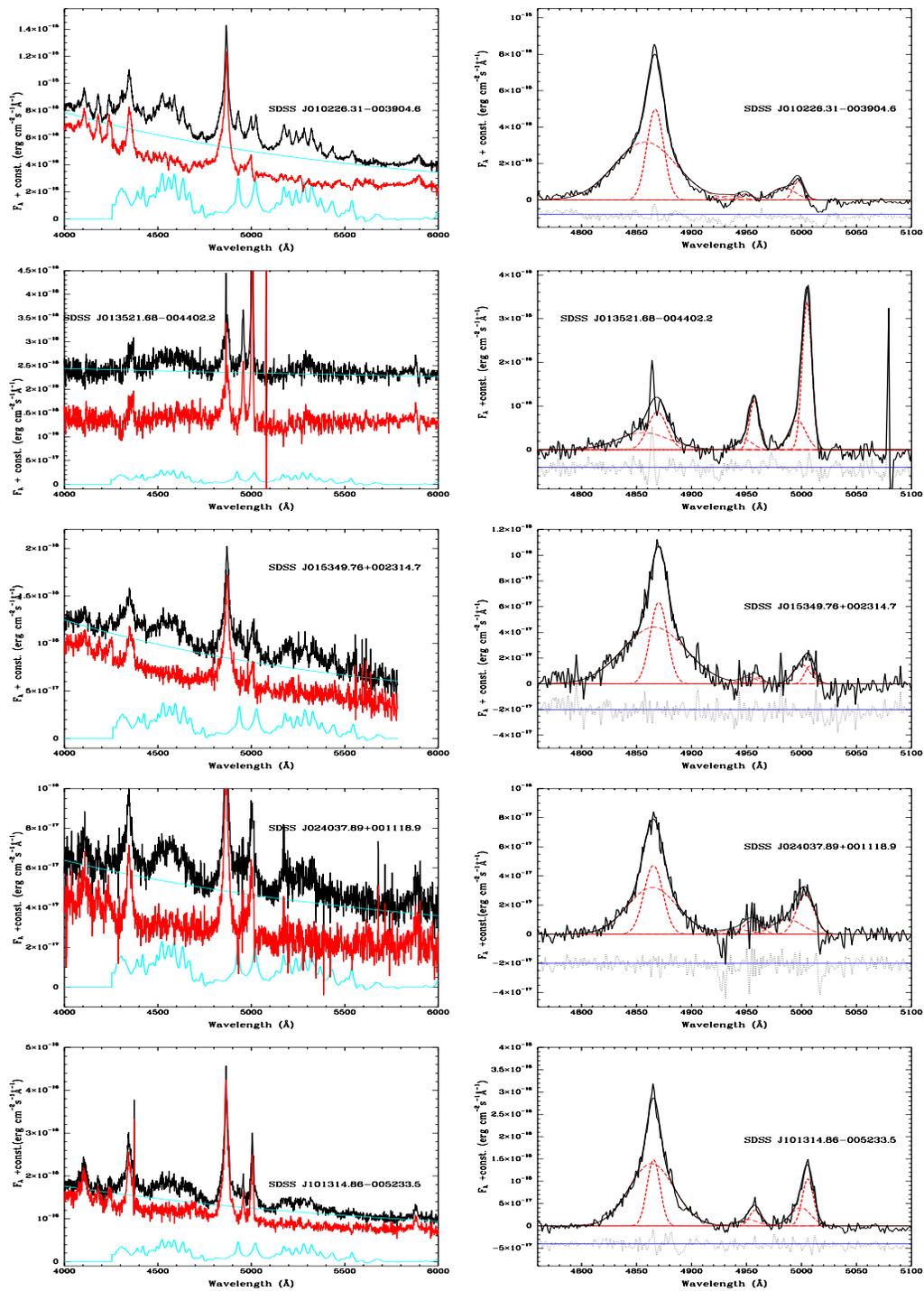,width=17cm,height=25cm}
\vglue-2.5cm \caption{Rest-frame spectra of ``blue outliers"
(left): the observed spectrum and a power-law continuum (top
curve), Fe II-subtracted spectrum (middle curve), and Fe II
spectrum (buttom curve); Multi-component fitting of the H$\beta$
and [O III]$\lambda\lambda$4959, 5007 (right): modeled composite
profile (thick solid line), individual components (the dotted
lines), the residual spectrum (lower panel).}
\end{center}
\end{figure*}

\newpage
\setcounter{figure}{2}
\begin{figure*}
\begin{center}
\vglue-1.5cm \epsfig{figure=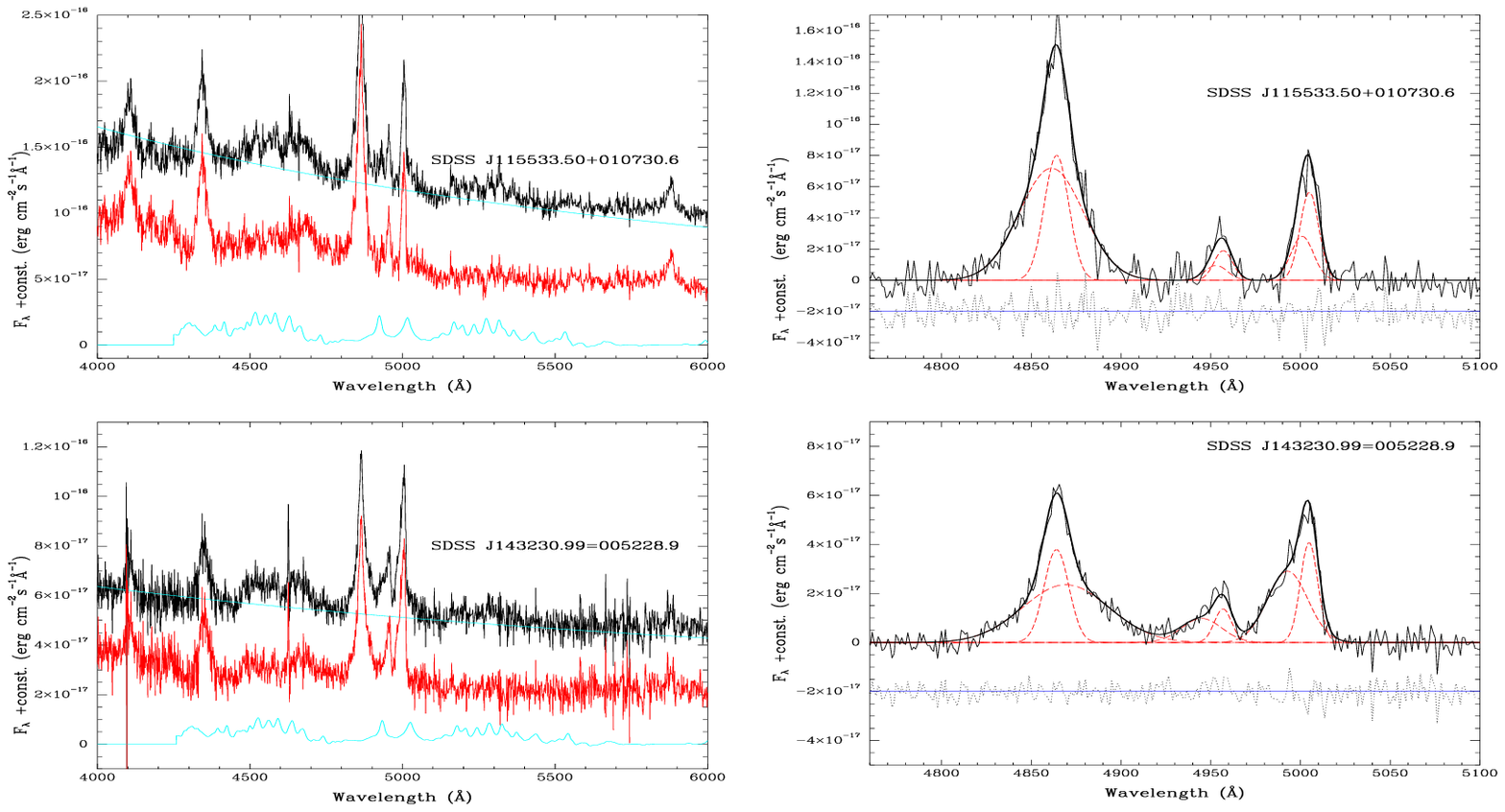,width=17cm,height=25cm}
\vglue-2.5cm \caption{Continued.}
\end{center}
\end{figure*}

\begin{figure*}
\begin{center}
\epsfig{figure=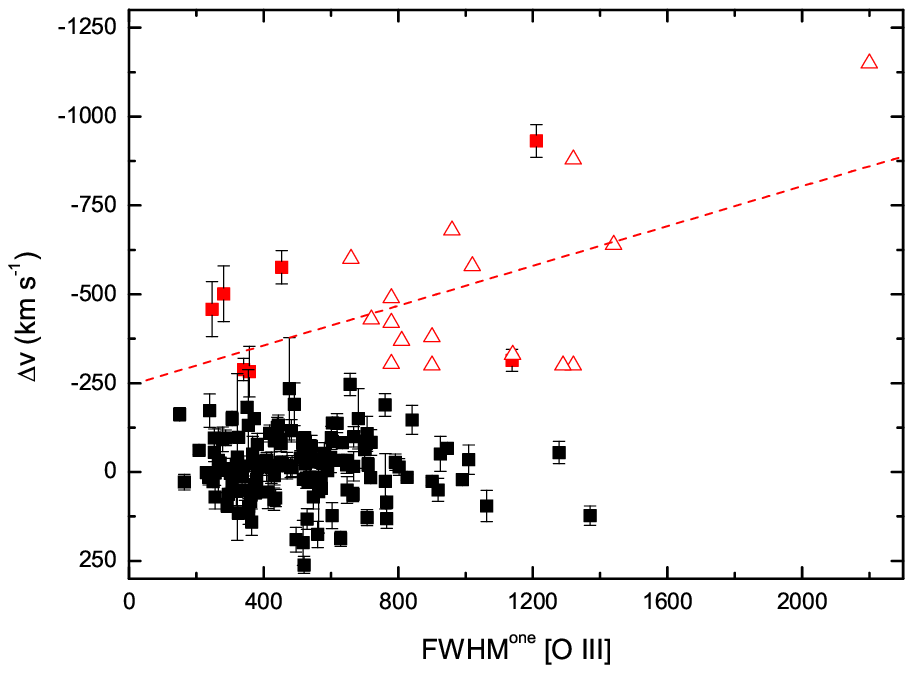,width=8cm}

\caption{The [O III] blueshift versus $FWHM^{one}([O III])$. Open
triangles denote the ``blue outliers" in the sample of
Aoki et al. (2005). The red dash line is our best fit for all
``blue outliers" from the sample of Aoki et al. (2005) and our SDSS
NLS1s sample.}
\end{center}
\end{figure*}

\end{document}